\documentclass[aps,pre,superscriptaddress,amsmath,twocolumn,showpacs,floatfix]{revtex4}
\usepackage{graphicx}

\begin{document}

\title{Probability distributions for polymer translocation}
\author{Cl\'ement Chatelain}
\affiliation{Department of Physics,
  Massachusetts Institute of Technology, Cambridge, MA 02139, USA}
\affiliation{Department of Physics, ENS Cachan, 61 Avenue du Pr\'esident Wilson, 94235 Cachan Cedex, France}
\author{Yacov Kantor}
\affiliation{Raymond and Beverly Sackler School of Physics and Astronomy, Tel Aviv University, Tel Aviv 69978, Israel}
\author{Mehran Kardar}
\affiliation{Department of Physics, Massachusetts Institute of Technology, Cambridge, MA 02139, USA}


\begin{abstract}

We study the passage (translocation) of a  self-avoiding polymer through 
a membrane pore in two dimensions. 
In particular, we numerically measure the probability distribution 
$Q(T)$ of the translocation time $T$, and the distribution $P(s,t)$ of 
the translocation coordinate $s$ at various times $t$. 
When scaled with the mean translocation time $\langle T\rangle$, 
$Q(T)$ becomes independent of polymer length, and decays
exponentially for large $T$.  
The probability  $P(s,t)$ is well described by a Gaussian at short times, 
with a variance that grows sub-diffusively as $t^{\alpha}$ with $\alpha\approx 0.8$. 
For times exceeding  $\langle T\rangle$, $P(s,t)$
of the polymers that have not yet finished their translocation has a 
non-trivial stable shape.

\end{abstract}

\maketitle

\section{Introduction}

Translocation of a long polymer through a narrow pore in a membrane
has been extensively studied experimentally during the last 
decade~\cite{kasian,akeson,mel}.   It is important in many biological 
and chemical processes such as viral injection of DNA into a host, RNA 
transport through nanopore of the nuclear membrane~\cite{bamolbel}. 
It may also have practical applications such as the possibility to
``read" a DNA or RNA sequence by passing it  through a nanopore 
such as microfabricated channels or  the $\alpha$-hemolysin 
channel~\cite{meller_review}.
Understanding the dynamics of translocation is also of inherent 
fundamental interest.
Theoretically, the short time  behavior has been investigated~\cite{mathe} 
by considering ever more detailed models of the interaction between 
the polymer and the pore in the membrane. 
The microscopic details should not be necessary to understand the scaling
of the passage time for very long polymers, where it should suffice to
resort to rather simple models of the polymer and  the membrane. 

It is convenient to track the process with  a single variable $s$, 
called the  {\em translocation coordinate}, that is the
monomer number at the pore~\cite{muthu,Lubensky,park,chern}, and also 
indicates how much of the polymer has passed to the other side.
In terms of this variable, the translocation process begins when
the first monomer enters the pore ($s=1$) and ends when the last
monomer exits to the other side at $s=N$,  at the translocation time $T$.
If the  translocation process is sufficiently slow (to allow for the equilibration
of the polymer), the mean force acting on the monomer at the hole can be 
obtained from a  simple calculation of free-energies.
This is a reasonable approximation for the experimental results
of relatively short polymers~\cite{meller_review}. 
The reduction of entropy creates a weak potential barrier, and the
translocation problem becomes equivalent to the escape of a 
`particle' (the translocation coordinate) over this barrier. 
As is the case for diffusion over an interval of length $N$, the mean
translocation time is found to scale as  $\langle T\rangle\sim N^2$.
(The logarithmic potential due to entropy is too weak to modify this scaling.)

In the absence of hydrodynamic interactions, 
the relaxation time $\tau$ of a polymer scales with the number of monomers
as $N^{1+2\nu}$~\cite{deGennes_book,Doi}, where the
exponent $\nu$ characterizes scaling of the radius of gyration $R_g$ of the polymer 
by $R_g\sim N^\nu$. In good solvent $\nu={3/4}$ in
two dimensions (2D), and $\nu\approx0.59$ in three dimensions (3D).
Note that $\tau$  is of the order of the time the polymer needs to diffuse its own $R_g$. 
Since $\tau$ grows faster than $N^2$,  the quasi-equilibrium approach
to translocation, described in the previous paragraph, must fail for 
sufficiently large $N$.
The relaxation process slows down the passage of the polymer, and the stochastic
forces acting on the translocation coordinate must be anti-correlated.
Initial numerical simulations suggest~\cite{ckk}  that the resulting mean
translocation time $\langle T\rangle$ scales like the 
relaxation time $\tau$, i.e. $\langle T\rangle\sim N^{1+2\nu}$.
This suggests that, to the extent that the resulting process can be regarded as stationary,
the translocation coordinate $s$ executes anomalous (subdiffusive) motion.
Simple scaling considerations lead to the
conclusion~\cite{ckk}, that the variance of the translocation
coordinate $\delta s^2$ increases with time $t$ as $t^\alpha$ with 
$\alpha={2/(1+2\nu)}$. This power is obtained by the requirement 
that for $t=\langle T\rangle$,
$\delta s^2\sim N^2$, i.e. the translocation is complete.

Anomalous diffusion in translocation is closely related to
the behavior of a tagged monomer in a long polymer~\cite{KBG}: 
At short time scales, before a monomer feels the effects of its neighbors,
it undergoes rapid normal diffusion with  diffusion constant $D_o$.
At very long times, exceeding the relaxation time $\tau$, 
the entire polymer (and hence each monomer)
diffuses along with the center of mass of the polymer, with a slow
diffusion constant of $D_o/N$.
At intermediate times, the fluctuations of the monomer are independent of
the total length, and to match the final $N$-dependent diffusion constant
must be characterized by anomalous diffusion~\cite{KBG}.
Both the tagged monomer and the translocation coordinate are slowed
down by the couplings to the rest of the polymer, and undergo
{\em subdiffusion} since  the variances 
of the relevant variables increase sublinearly ($\alpha<1$).
There are of course important differences: the diffusion of
a tagged monomer is executed in $d$-dimensional real space, while 
the translocation coordinate moves along the one dimensional axis
of monomer numbers. The scaling exponents are also different, 
although their values are derived from closely related considerations. 
Recently we studied the constrained motion of a tagged monomer
as an indirect means of gaining insight into the distribution of
translocation times~\cite{KantKard1}.

There has been much recent progress in the
theoretical modelling of  translocation: hydrodynamic 
interactions were taken into account~\cite{storm,tian}, and an
intuitive scaling picture of polymer translocation under the 
influence of a force~\cite{GrosNech,sakaue} was developed. A 
variety of scaling regimes with force applied to the end-point or at the 
pore have been investigated numerically in some detail~\cite{luo}.
Some recent studies~\cite{panja,dubb} suggest that the translocation
process in 3D maybe even slower than dictated by the relaxation time.
If so, this would weaken the analogy between translocation and
the motion of a tagged monomer.  (The accuracy of these claims 
is questioned in further work~\cite{luocom}.)

Since the translocation process is terminated when $s$ reaches one
end of the polymer, one may draw an analogy to
the anomalous diffuser in the presence of absorbing 
boundaries. One approach frequently used to describe subdiffusion
is the fractional diffusion equation (FDE)~\cite{mkreview}.
Solutions of FDE in the presence of absorbing boundaries predict that
for large $t$ the absorption probability $Q(t)$ decays as 
$1/t^{\alpha+1}$~\cite{fpinf}. In the case of subdiffusion ($\alpha<1$) 
this decay is so slow that the mean absorbtion time diverges. 
By applying this analogy to translocation it has been suggested~\cite{lua} 
that the mean translocation time is infinite, and there is a 
numerical study~\cite{dubb} lends support for a power-law tail in the distribution 
of translocation times.
If so, this would imply that the experimentally and numerically
measured translocation times are artifacts of the finite duration of
the experiment. However, this proposition is not supported by 
experiments or other numerical simulations.

To address this controversy, we recently considered~\cite{KantKard1} 
the motion of a tagged monomer belonging to a very long phantom 
(Gaussian) polymer, moving in one-dimension between two absorbing boundaries.
We demonstrated that at least in this case, $Q(t)$ decays exponentially even though
the monomer undergoes subdiffusion. 
While this casts strong doubts to the generality and relevance of the 
conclusions based on FDE, it does not directly address the dynamics of
translocation, and thus not necessarily contradict the  conclusions of 
Ref.~\cite{dubb}.
Thus, in this work, we perform a direct and detailed study of the translocation
process for a self-avoiding polymer in 2D. 
We concentrate on the behavior of the distribution of the translocation times, 
and also on the stationary distribution of the translocation coordinate
at very long times. In Sec.~\ref{sec:model}
we describe our numerical model and the Monte Carlo (MC) procedure.
The results presented in Sec.~\ref{sec:results} demonstrate that for large
$t$ the distribution of the translocation times decays exponentially,
while the long-time distribution of the translocation coordinate takes
a non-trivial form.

\section{The model and simulations}\label{sec:model}

Simulations of the self-avoiding polymer translocating through a membrane
were performed with a fluctuating bond polymer model~\cite{carm} in 2D.
In this model, the $N$ monomers are restricted to the sites of a square lattice.
Excluded volume interactions are implemented by forbidding two monomers to
be closer than 2 lattice constants, while the polymeric character is enforced
by requiring the separation between monomers adjacent along the chain to
be less than $\sqrt{10}$ lattice constants.
This choice of minimal and maximal distances ensures that
the polymer never intersects itself. 
The model contains no energy scale, leading to an extremely simple Monte Carlo
procedure: An elementary move consists of an attempt to move a randomly
selected monomer by one lattice spacing in an arbitrarily chosen direction.
If the new configuration is permitted, the step is executed;
otherwise, the configuration remains unchanged. 
One MC time unit is composed of $N$  elementary moves. 
This model closely resembles tethered spheres used in continuum~\cite{kkn} simulations. 
We previously used this model to demonstrate the anomalous dynamics
of polymer translocation~\cite{ckk}. The membrane intervening membrane
in simulations has a  thickness of two lattice constants, with a hole that is 
three lattice spacings wide. The tight size ensures that only one monomer can pass 
through the hole, and enables a unique designation of the monomer $s$ 
which separates the polymer segments on the two sides of the membrane. 

The translocation process in actuality involves several complicating factors:
The polymer located on one side of the membrane must first reach the pore
such that one end enters the pore.
In the absence of strong driving force, it is then quite likely that the polymer
retracts and does not pass through to the other side until a number of
such attempts. Both of these processes have been discussed in the literature. 
Since we are only interesting in the anomalous dynamics during the translocation process,
we implement a computation procedure that is different from the usual experimental conditions. 
In our simulations, an the initial configuration is constructed by fixing the monomer $s=N/2$
in the hole, and equilibrating the remaining monomers for more than the relaxation time
(which is proportional to $N^{2.5}$)~\cite{deGennes_book}. 
After this equilibration is finished, at time $t=0$, the fixed monomer is allowed 
to move freely. The simulation ends at time $t=T$ when the entire 
polymer is on {\em either side} of the membrane. We denote $T$ the 
{\em translocation time}. 
The procedure is repeated a large number of times for each polymer size $N$,
to construct the probability $Q_N(T)$.
For each simulation run, we also record the trajectory $s(t)$ of the translocation coordinate. 
Consequently, we are able to monitor the evolution of the distribution $P(s,t)$. 
At the starting moment $P(s,0)=\delta_{s,N/2}$, and it subsequently broadens as
the time increases. It should be noted that as $t$ reaches typical
translocation times, the fraction of polymers that has completed the process starts to
grow, and consequently the probability distribution $P(s,t)$, which is normalized
for each $t$, is obtained from a decreasing sample of runs. 
Another drawback is that the simulation times increase as $N^{3.5}$, making it 
difficult to obtain good statistics in the interesting limit of large $N$.
We performed simulations for $N=8,~16,\dots,~128$, and 256. 
Most of the results presented in the paper
correspond to $N=128$, for which we are able to obtain sufficiently many
samples.

\section{Results}\label{sec:results}

\begin{figure}[t]
\includegraphics[width=0.48\textwidth]{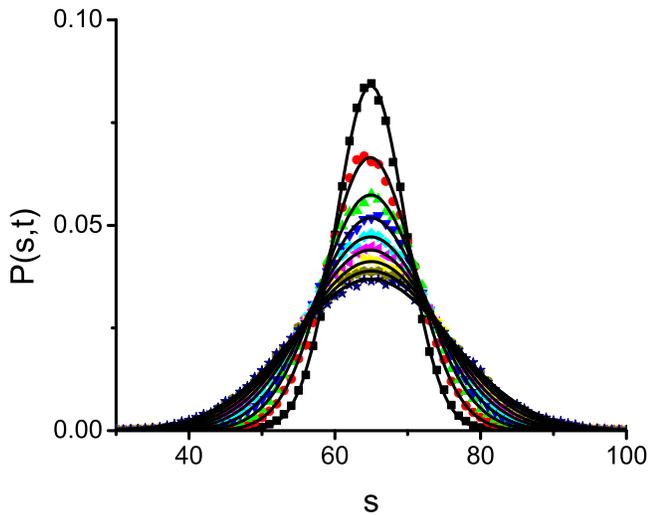}
\caption{(Color online) Probability distribution of the translocation coordinate $s$, of a
polymer with $N=128$ monomers for MC times $t=10^4,~2\times 10^4,\cdots,$
and $9\times 10^4$ 
(from narrowest to the widest distribution), which are significantly shorter
than the mean translocation time. The results are obtained from 10,000
independent runs. Continuous lines represent
Gaussian fits to these distributions.}
\label{fig:pdfs-gaussian}
\end{figure}

Each of our simulations begins at $s=N/2$, and as the time $t$ increases 
the distribution $P(s,t)$ becomes broader. As long as $t$ is significantly 
shorter than the mean translocation time $\langle T\rangle$,  the 
distribution of $s$ resembles a Gaussian, as can be seen in
Fig.~\ref{fig:pdfs-gaussian}, and is very different from the shapes
obtained from the solutions of FDE.  The plots depict the behavior of $s$
for $N=128$, where $\langle T\rangle\approx 2.9\times 10^6$ MC time units.
Similar shapes were observed in the simulations
of a tagged monomer in a phantom polymer (see Fig.~1 in Ref.~\cite{KantKard1}), 
where one can {\em prove} that the distributions are indeed Gaussian. 
In our case, there is no analytical proof, and we instead numerically examined the 
fourth cumulant $\kappa_4$ of the distribution, which vanishes for a Gaussian PDF. 
Since the variable $s$ is discrete, $\kappa_4$ does not vanish, but should become 
significantly smaller than the squared second cumulant 
(which is the variance of the distribution $\delta s^2$), as the width 
of the distribution increases. We find that once this width
exceeds 2, the cumulant $\kappa_4$ becomes only a few percent of
$(\delta s^2)^2$, and the deviation from zero is probably caused
by the statistical errors. Thus within our statistical accuracy, the
PDF at short times is indistinguishable from a Gaussian.

\begin{figure}[t]
\includegraphics[width=0.48\textwidth]{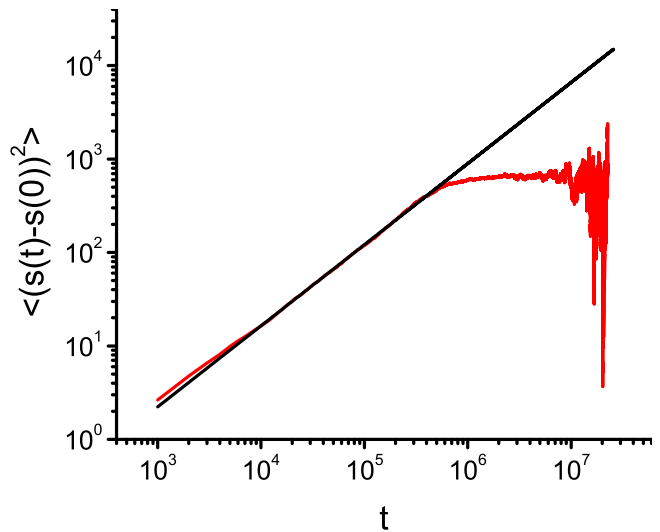}
\caption{(Color online) The mean squared displacement of 
the translocation coordinate $s$ as a function of MC time $t$, obtained from
$10^4$ runs of a 128-monomer polymer. 
The straight line represents the power-law fit $t^\alpha$ in the interval
$10^4<t<1.2\times 10^5$ with exponent $\alpha=0.86$.}
\label{fig:deltas2}
\end{figure}

As the time goes on, the variance $\delta s^2$ is expected to increase 
as $t^\alpha$, with $\alpha\approx 2/(1+2\nu)=0.8$, and eventually saturate at 
values of order $N^2$. Figure~\ref{fig:deltas2} depicts such dependence 
on a logarithmic scale. For a range of times longer than $10^4$ this line has 
a straight segment with slope 0.86, which is slightly larger than the 
above value, and is consistent with the other numbers quoted in the 
literature~\cite{ckk,luo}.  The statistical accuracy
of the calculated exponent is better than the last significant digit
of the number. However, we believe that systematic errors related
to crossovers and specific choice of the fitting range introduce
significantly larger (few percent) errors. 
Curves for various values of $N$
saturate approximately at $\delta s^2\approx (N/5)^2$,
and corresponding saturation  time $T_1$ is obtained by extrapolating the (low-$t$) power-law behavior to
this value. For the various lengths $N$ used in our simulations 
the ratio $\langle T\rangle/T_1$ is approximately constant.

\begin{figure}[t]
\includegraphics[width=0.48\textwidth]{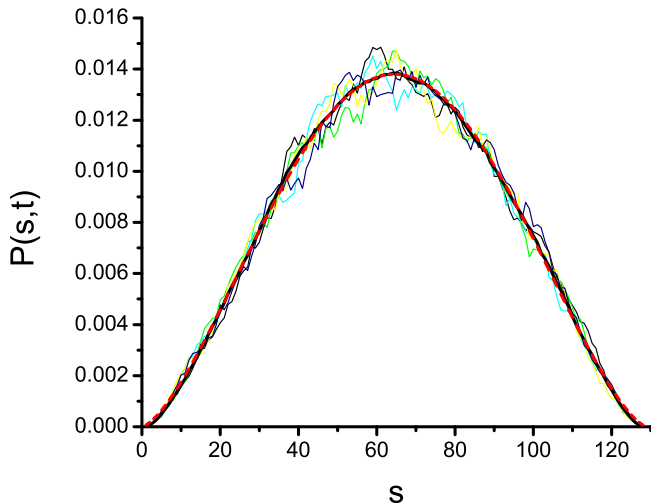}
\caption{(Color online) Probability distribution of the translocation variable $s$
of the subset of polymers  with $N=128$ monomers that did not complete translocation at MC times
$t=40\times 10^5,~42\times 10^5,\cdots$, and $48\times10^5$ (thin lines), 
which exceed the mean translocation time. These graphs were obtained by performing $10^4$ independent
runs out of which only 15--20\% survive to the times when the data is collected.
The thick solid line is the average of 10 graphs in the  range 
$4\times 10^6\le t\le4.9\times 10^6$. 
The dashed line depicts the fit function $A\sin^k(s\pi /(N+1))$ with $k=1.44$.}
\label{fig:pdfs-assymp}
\end{figure}

As the simulation time $t$ exceeds $\langle T\rangle$, a significant fraction of
polymers complete their translocation process. For the subsets of runs that survive into
such long times, the distribution of $P(s,t)$ reaches a stable shape. In the
case of normal diffusion between two absorbing boundaries, the limiting shape is a
sine-function that vanishes linearly near the boundaries. 
This shape reflects the lowest eigenfunction of the diffusion operator (Laplacian), 
corresponding to the longest decay time (eigenvalue). 
Since we do not know of a corresponding differential equation for the translocation 
coordinate, we do not know the corresponding limiting shape. 
Some insight into possible solutions can be gleaned by considering a fractional
diffusion operator, as in the case of the Laplacian raised to the power $1/\alpha$,
with absorbing boundary conditions (see, e.g.  Zoia {\em et al.}~\cite{zoia}).
In the absence of absorbing boundaries the squared width of the distribution produced 
by such a fractional Laplacian increases as $t^\alpha$. 
In the presence of the boundaries, the eigenstates are not known for general $\alpha$. 
However, it is known that near the absorbing boundaries the eigenfunction goes to zero 
nonlinearly, with an exponent of  $k=1/\alpha$. 
For normal diffusion, this naturally reduces to the expected linear form.
For subdiffusion, the eigenvalues of this operator vanish faster than linearly;
in the case $\alpha=0.8$ with exponent $k=1.25$. 
Figure~\ref{fig:pdfs-assymp} depicts 
$P(s,t)$ for $N=128$ and for times exceeding $\langle T\rangle$. The 
statistical accuracy of these results is not very good, since a significant 
fraction of the polymers have already translocated. The accuracy is particularly 
poor near the endpoints of the graph where the probability approaches zero. 
Nevertheless, we observe that the function seems to decay faster
than linearly and slower than quadratically. The overall shape of the distribution
can be approximated by  the function $A\sin^k(s\pi /(N+1))$. We find a good fit with $k=1.44$. 
Qualitatively, Fig.~\ref{fig:pdfs-assymp} resembles the results obtained
for tagged monomer diffusion (see Fig.~8 in Ref.~\cite{KantKard1}),
but with a different exponent $k$.

\begin{figure}[t]
\includegraphics[width=0.45\textwidth]{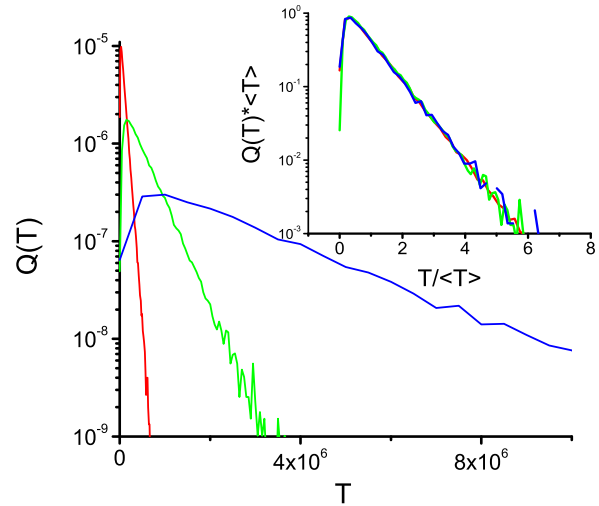}
\caption{(Color online) Probability distribution of translocation 
time $T$ for  $N=32$, $64$ and $128$ (left to right), obtained from 
100,000, 78,600 and 8,600 runs, respectively. 
The inset demonstrates the collapse of the probabilities when $T$ is
scaled with its average value $\langle T\rangle$.}
\label{fig:pdft}
\end{figure}

We also studied the probability distribution $Q(T)$ of the translocation time. In the range from
$N=8$ to $N=256$ we find that the mean translocation time $\langle T\rangle$
increases as $N^{2.51}$ consistent with $1+2\nu=2.5$, and in accord 
with previous work~\cite{luo,ckk}. As in the case of the
numerical estimate of $\alpha$, the statistical errors are smaller than the last
significant digit, but we should beware of systematic errors. 
For example, the exponent 2.51 should correspond to $\alpha\approx0.80$, 
which is smaller than the directly measured value of 0.86, 
and indicates the importance of systematic errors. 
Figure~\ref{fig:pdft} depicts $Q(T)$ for three values of $N$ on a semi-logarithmic
scale. We clearly see an exponential decay $Q(T)\sim \exp(-T/T_0)$ for 
large $T$ in each of the graphs. The decay constant $T_0$ increases with increasing
$N$. The ratio $\langle T\rangle/T_0$ is approximately a constant. Moreover,
in terms of rescaled times $T'=T/\langle T\rangle$ the distribution
becomes independent of $N$, as can be seen in the inset in Fig.~\ref{fig:pdft}.

\section{Discussion and conclusion}\label{sec:discuss}

In this work we performed a detailed study of distribution functions associated
with the translocation of  a 2D model of a self-avoiding polymers. 
Our results clearly indicated an exponential decay of the PDF $Q(T)$ 
for large translocation times $T$, and thus  exclude power-law~\cite{dubb}
or stretched exponential~\cite{NOB} behavior.
The distribution of the translocation variable $s$ both at short and
long times exhibits the behavior resembling that of a tagged 
monomer~\cite{KantKard1}. 
There is some similarity in the behavior of the long-times stationary
distributions from our simulations, and solutions of a fractional Laplacian
with absorbing boundaries~\cite{zoia}. However, the accuracy of our results precludes
definitive statements regarding these long-time distributions.

Strong crossover effects are present in translocation for surprisingly
high values of $N$. In fact, the values of various exponents reported in
the literature differ beyond their nominal error bars. One may hope
that for $N$ as large as 1,000 this difficulty can be overcome. Unfortunately,
accumulating very large statistics for such large $N$ is currently beyond
our ability.

\begin{acknowledgments}
This work was supported by the National Science Foundation Grant No. 
DMR-04-26677 (M.K.) and
by the Israel Science Foundation (Y.K.). C.Ch. thanks the MIT-France Program.
\end{acknowledgments}


\begin{thebibliography}{25}

\bibitem{kasian} J. J. Kasianowicz, E. Brandin, D. Branton, and D. W. Deamer,
Proc. Natl. Acad. Sci. U.S.A.  {\bf 93}, 13770 (1996).

\bibitem{akeson}M. Akeson, D. Branton, J. J. Kasianowicz, E. Brandin, and D.
W. Deamer, Biophys. J. {\bf 77}, 3227 (1999).

\bibitem{mel}A. Meller, L. Nivon, E. Brandin, J. Golovchenko, and D. Branton,
Proc. Natl. Acad. Sci. U.S.A. {\bf 97}, 1079 (2000);
A. Meller and D. Branton, Electrophoresis {\bf 23}, 2583 (2002);
A. Meller, L. Nivon, and D. Branton, \prl {\bf 86}, 3435 (2001).

\bibitem{bamolbel} B. Albert, {\it Molecular Biology of the Cell} (Garland, New York, 1994).

\bibitem{meller_review} A. Meller, J. Phys: Cond. Matter. {\bf 15}, R581 (2003).

\bibitem{mathe}J. Mathe, A. Aksimentiev, D. R. Nelson, K. Schulten, and A.
Meller, Proc. Natl. Acad. Sci. U.S.A. {\bf 102}, 12377 (2005);
A. Aksimentiev, J. B. Heng, G. Timp nad  K. Schulten,
Biophys. J. {\bf 87}, 2086 (2004).

\bibitem{muthu} M. Muthukumar, \jcp {\bf 111}, 10371 (1999).

\bibitem{Lubensky} D. K. Lubensky and D. R. Nelson, Biophys. J.
{\bf 77}, 1824 (1999).

\bibitem{park}W. Sung and P. J. Park, \prl {\bf 77},  783 (1996);
P. J. Park and W. Sung, \jcp {\bf 108},  3013 (1998).

\bibitem{chern}Sh.-Sh. Chern, A. E. C\'ardenas, R. D. Coalson, 
\jcp {\bf 115}, 7772 (2001).

\bibitem{deGennes_book}P.-G. de Gennes, {\it Scaling Concepts in
Polymer Physics}, (Cornell Univ. Press, Ithaca, NY, 1979).
 
\bibitem{Doi}M. Doi and S. F. Edwards, {\it The Theory of Polymer
Dynamics} (Clarendon Press, Oxford, 1986).

\bibitem{ckk}J. Chuang, Y. Kantor, and M. Kardar, \pre {\bf 65}, 011802 (2001);
Y. Kantor, and M. Kardar, \pre {\bf 69}, 021806 (2004).
 
\bibitem{KBG}K. Kremer and K. Binder, \jcp {\bf 81}, 6381
(1984); G.S. Grest and K. Kremer, \pra {\bf 33}, 3628 (1986).

\bibitem{KantKard1} Y. Kantor and M. Kardar, \pre {\bf 76}, 061121 (2007).
 
\bibitem{storm}A. J. Storm, C. Storm, J. Chen, H. Zandbergen, J. F. Joanny,
and C. Dekker, Nano Lett. {\bf 5}, 1193 (2005).

\bibitem{tian}P. Tian and G. D. Smith, \jcp {\bf 119}, 11475 (2003).

\bibitem{GrosNech}A. Yu. Grosberg, S. Nechaev, M. Tamm, O. Vasilyev, 
\prl {\bf 96}, 228105 (2006).

\bibitem{sakaue}T. Sakaue, \pre {\bf 76}, 021803 (2007).

\bibitem{luo}K. Luo, I. Huopaniemi, T. Ala-Nissila, S.-Ch. Ying, 
\jcp {\bf 124}, 114704 (2006); 
K. Luo, T. Ala-Nissila, S.-Ch. Ying, \jcp {\bf 124}, 034714 (2006);
I. Huopaniemi, K. Luo, T. Ala-Nissila, S.-Ch. Ying, \jcp {\bf 125}, 124901 (2006);
I. Huopaniemi, K. Luo, T. Ala-Nissila, S.-Ch. Ying, \pre {\bf 75}, 061912 (2007).
 
\bibitem{panja}J. K. Wolterink, G. T. Barkema, and D. Panja, 
\prl {\bf 96}, 208301 (2006); 
D. Panja, G. T. Barkema and R. C. Ball, arXiv:cond-mat/0610671;
and J. Phys.: Cond. Matt. {\bf 19}, 432202 (2007).

\bibitem{dubb}J. L. A. Dubbeldam, A. Milchev, V. G. Rostiashvili, and
T. A. Vilgis, \pre {\bf 76}, 010801 (R) (2007).

\bibitem{luocom}K. Luo, T. Ala-Nissila, S.-Ch. Ying, P. Pomorski and
M. Kattunen, arXiv:0709.4615.

\bibitem{mkreview}R. Metzler and J. Klafter, Phys. Rep. {\bf 339}, 1 (2000);
and J. Phys. A: Math. Gen. {\bf 37}, R161 (2004).

\bibitem{fpinf}S. B. Yuste, K. Lindenberg, \pre {\bf 69}, 033101 (2004);
 M. Gitterman, \pre {\bf 62}, 6065 (2000); and \pre {\bf 69}, 033102 (2004).
 
\bibitem{lua}R. C. Lua, A. Y. Grosberg, \pre {\bf 72}, 61918 (2005).

\bibitem{carm}I. Carmesin, K. Kremer, Macromol. {\bf 21}, 2819 (1988).

\bibitem{kkn}Y. Kantor, M. Kardar and D. R. Nelson, \prl {\bf 57} (1986);
and  \pra {\bf 35}, 3056 (1987).

\bibitem{zoia} A. Zoia, A. Rosso, and M. Kardar, \pre {\bf 76}, 021116 (2007).

\bibitem{NOB}S. Nechaev, G. O. Oshanin and A. Blumen, J. Stat. Phys. {\bf 98},
281 (2000).

\end{thebibliography}
\end{document}